\RequirePackage[undo-recent-deprecations]{expl3}
\documentclass[a4paper,fleqn]{cas-sc}
\usepackage{soul}
\usepackage[sort&compress,numbers]{natbib}
\usepackage{lineno}
\usepackage{setspace}

\usepackage{cuted}

\def\tsc#1{\csdef{#1}{\textsc{\lowercase{#1}}\xspace}}
\tsc{WGM}
\tsc{QE}
\tsc{EP}
\tsc{PMS}
\tsc{BEC}
\tsc{DE}

\begin{document}
\let\WriteBookmarks\relax
\def\floatpagepagefraction{1}
\def\textpagefraction{.001}
\shorttitle{Polaron Transport in Porous Graphene Nanoribbons}
\shortauthors{Cunha \textit{et~al}.}

\title [mode = title]{Charge Density Wave Transport in Porous Graphene Nanoribbons}

\author[1]{Wiliam F. da Cunha}
\author[1]{Marcelo L. Pereira J\'unior}
\author[2]{William F. Giozza}
\author[2]{Rafael T. de Sousa Junior}
\author[1,3]{Luiz A. Ribeiro J\'unior}
\cormark[1]
\ead{ribeirojr@unb.br}
\author[1]{Geraldo M. e Silva}

\address[1]{Institute of Physics, University of Bras\'ilia, Bras\'ilia, Brazil.}
\address[2]{Department of Electrical Engineering, University of Bras\'{i}lia 70919-970, Brazil.}
\address[3]{PPGCIMA, Campus Planaltina, University of Bras\'{i}lia, 73345-010, Bras\'{i}lia, Brazil.}


\begin{abstract}
Porous graphene (PG) forms a class of graphene-related materials with nanoporous architectures. Their unique atomic arrangements present interconnected networks with high surface area and high pore volume. Some remarkable properties of PG, such as high mechanical strength and good thermal stability, have been widely studied. However, their electrical conductivity, and most importantly, their charge transport mechanism are still not fully understood. Herein, we employed a numerical approach based on a 2D tight-binding model Hamiltonian to first reveal the nature of the charge transport mechanism in PG nanoribbons. Results showed that the charge transport in these materials is mediated by charge density waves. These carrier species are dynamically stable and present very shallow lattice distortions. The porosity allows for an alternative to the usual arising of polaron-like charge carriers and it can preserve the PG semiconducting character even in broader nanoribbons. The charge density waves move in PG within the optical regime with terminal velocities varying from 0.50 up to 1.15 \r{A}/fs. These velocities are lower than the ones for polarons in conventional graphene nanoribbons (2.2-5.1 \r{A}/fs).
\end{abstract}


\maketitle
\doublespacing
      
\section{Introduction}
\label{sec1}

One of the most important goals of material engineering field is to conceive systems that combine the advantages of different materials for a given application \cite{yang2012towards,allison2006integrated,hubbell2009materials,simske1997porous}. For instance, the optical and mechanical properties of graphene sheets are very well established and celebrated \cite{geim2010rise,novoselov2004electric}. However, this system lacks at least one crucial feature for its usage in optoelectronic devices, namely a finite bandgap \cite{novoselov2011nobel,novoselov2007electronic,neto2009electronic}. The atomic-level control for structural modifications in the graphene sheet is arguably an option to give rise to related materials that present such desirable properties \cite{kotakoski2014imaging,soler2016engineering,lahiri2010extended,robertson2012spatial}. The search for systems based on graphene is the rationale behind the great effort in investigating materials of different symmetries \cite{ferrari2015science,tang2013graphene,bonaccorso2015graphene}. As graphene, all those systems are solely composed of carbon atoms. Their distinct configurations are reflected in very different behavior as far as electronic properties are concerned \cite{singh2011graphene,li2008graphene,huang2011graphene}. Among them, Porous Graphene (PG) has been successfully synthesized \cite{porous,jiang2014design} and presented a direct band gap in the range of 2.3-3.2 eV \cite{bandporous1,jiang2014design}, which is of the order of what is desired for an organic semiconductor device \cite{simske1997porous,jiang2009porous,yan2012advanced,celebi2014ultimate,koenig2012selective}.


Very often, polarons and bipolarons are responsible for the charge transport in graphene-based materials \cite{da2016polaron,silva2019geraldo,abreu2019stability}. They are a product of the energetic interplay between the electronic and the lattice degrees of freedom. An alternative mechanism for charge transport that can be mainly found in linear chains is charge density waves \cite{cdw}. It consists of a modulation of the charge density corresponding to the conduction electrons, which is accordingly accompanied by a related periodic distortion of the lattice. Be that as it may, external excitation can give rise to the appearance of oppositely charged bound states known as excitons in either scenario \cite{yang2007excitonic}.

A usual procedure to obtain non-vanishing bandgap materials from graphene is to cut the sheet in narrow stripes of different widths and edges, thus yielding Nanoribbons \cite{cai2010atomically,son2006energy,son2006half,han2007energy}. The confinement effects of the wave function of the charge carrier in one dimension contribute to the arising of desirable properties from the mobility point of view \cite{han2010electron,wakabayashi2009electronic,martin2012electronic,baringhaus2014exceptional,li2008role}. The literature has dedicated some effort in the theoretical and experimental description of PG conducting properties in recent years \cite{junior2020thermomechanical,xiao2011hierarchically,tao2014tunable,hankel2012asymmetrically,xu2012porous,mukherjee2014defect,brunetto2012nonzero,singh2016strategic,lu2016flexible,lu2016flexible,surwade2015water,du2010multifunctional,schrier2010helium}. Nonetheless, an investigation of the dynamics of the charge carriers in PG nanoribbon is still absent, even being crucial to the understanding of the transport properties of this system.

In this work, we studied the transport of charge carriers in PG nanoribbons by employing a two-dimensional lattice relaxation endowed Hamiltonian. To understand the time-dependent properties of the system, we conducted a time evolution of the degrees of freedom of both electrons and lattice parts of the system. The crucial issue addressed here concerns the very nature of the transport mechanism in PG nanoribbons. Based on our numerical approach, the findings revealed that charge density waves (more common in low dimensional systems) mediated the charge transport mechanism in PG nanoribbons. 

\section{Methodology}
\label{sec2}

\begin{figure*}[pos=ht]
	\centering
	\includegraphics[width=0.95\linewidth]{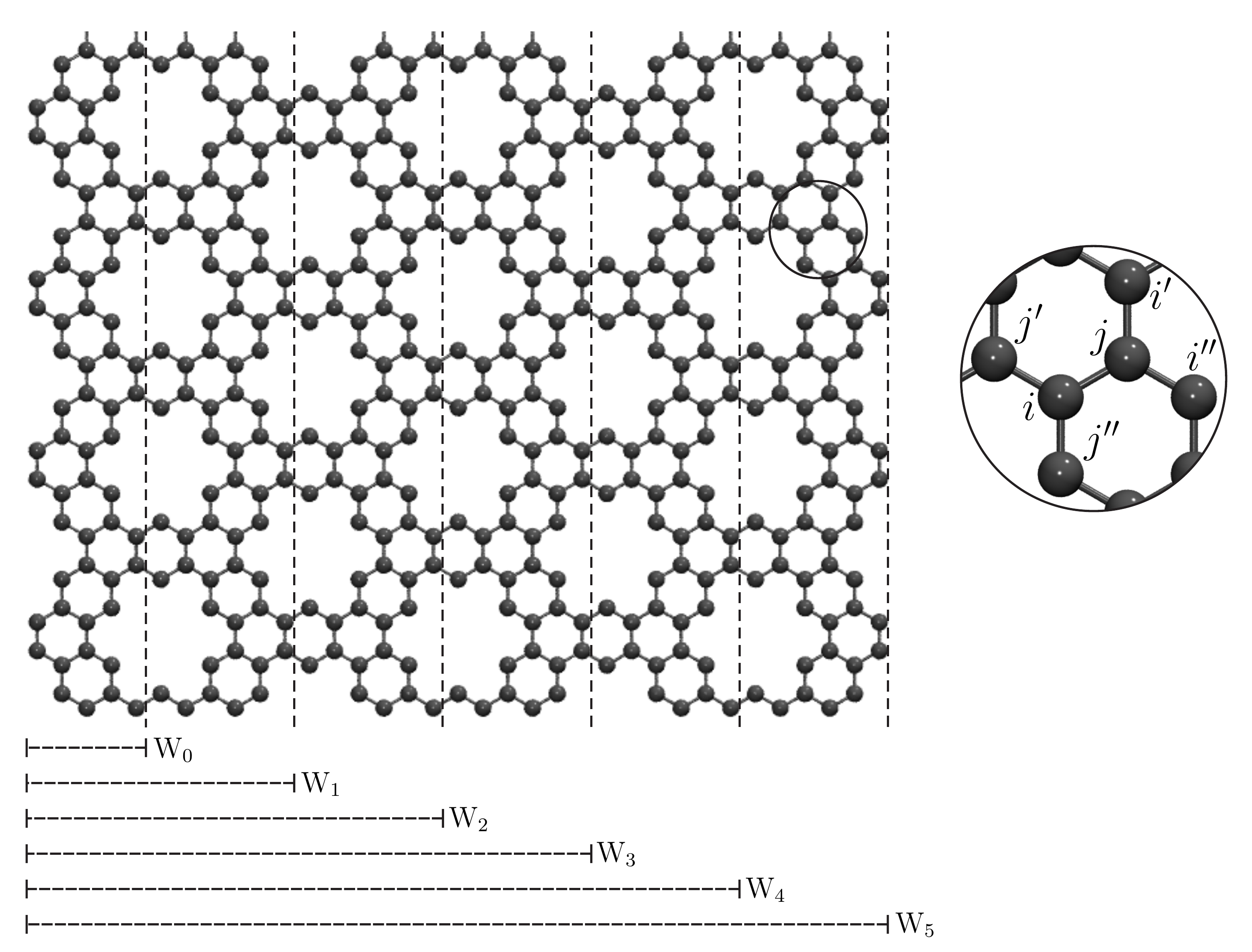}
	\caption{Schematic representation of the PG nanoribbons studied in this work. Each dashed vertical line illustrates a nanoribbon of different width. The inset contains the labeling of the sites used in our model Hamiltonian. Note that the dashed lines are just representing the limit for a particular ribbon width. In this sense, no dangling bonds are formed by cutting the lattice.}
	\label{fig:rep}
\end{figure*}

PG nanoribbons are obtained by periodically extracting carbon atoms from an armchair graphene nanoribbons. The left side of Figure \ref{fig:rep} represents a section of the resulting PG sheet, and the inset on the right shows the labeling of the sites to be used in the model. The dashed vertical lines represent PG nanoribbons of different widths. Note that these lines are just representing the limit for a particular ribbon width. In this sense, no dangling bonds are formed by cutting the lattice. The simulated systems have periodic boundary conditions in the vertical direction, as illustrated in Figure \ref{fig:rep}.

To correctly address the interplay between phonons and electrons, we resort to a hybrid model that combines classical and quantum treatments. In this sense, we applied the second quantization formalism of quantum mechanics to treat the electrons and the Euler-Lagrange equations to treat the lattice \cite{lima2006dynamical}. These two realms are connected by an electron-phonon coupling term that includes lattice relaxation to an otherwise two-dimensional tight-binding model \cite{da2019quasiparticle}. As is typical in $sp^2$ hybridized carbon systems, the displacement of the atoms in graphene is limited so that the electronic transfer integral for $\pi$-electrons can be expanded in first-order \cite{da2016polaron}, thus resulting in hopping terms such as
\begin{equation}
	\label{hop}
	t_{i,j} = t_0 - \alpha \eta_{i,j}.
\end{equation}
In the equation above, $t_0$ is the hopping integral of a regular lattice, $\alpha$ the electron-phonon coupling, and $\eta_{i,j}$ are the variations in the bond-lengths of two neighboring sites $i$ and $j$, as referred in the inset of Figure \ref{fig:rep}.

Considering the aforementioned expression for the hopping term, the resulting Hamiltonian is expressed as
\begin{equation}
\displaystyle 
		H = -\sum_{\langle i,j \rangle, s} \left(t_{i,j}^{\phantom{*}} C_{i,s}^\dag C_{j,s}^{\phantom{\dag}}+t_{i,j}^* C_{j,s}^\dag C_{i,s}^{\phantom{\dag}} \right ) + \frac12K\sum_{\langle i,j \rangle} \eta_{i,j}^2+\frac1{2M}\sum_ip_i^2,
\end{equation}
where $ \langle i,j \rangle $ means that the sum is performed over the neighboring sites labeled according to Figure \ref{fig:rep}; $C_{i,s}^{\phantom{\dag}}$ ($ C_{i,s}^{\dag}$) is the $\pi$-electron annihilation (creation) operator on site $i$ with spin $s$. The second term represents the harmonic modeling of the effective potential associated with $\sigma$-bonds between carbon atoms, with $K$ being the corresponding Hooke constant. As $p_i$ and $M$ are, respectively the atoms momentum and mass, the last term describes the kinetic energy associated to the sites. In a semi-empirical fashion we adopted the set of parameters most successfully reported in the literature: 2.7 eV for $t_0$, 21 eV/\AA$^2$ for $K$, and 4.1 eV/\r{A} for $\alpha$ \cite{barone2006electronic, de2012electron, ribeiro2015transport, kotov2012electron,yan2007electric, neto2009c, yan2007raman}. 

By beginning from a given initial set of coordinates $ \{\eta_{i,j}\} $, a self-consistent stationary solution (with $ p_i = 0 $) of the system is determined as follows: The ground state corresponding to the set $ \{\eta_{i,j}\} $ is obtained by diagonalizing the electronic Hamiltonian. The procedure also results in the determination of the eigenenergies of the system. From such states, it is possible to obtain the expected value of the system's Lagrangean. This operator is then used in Euler-Lagrange equations, and a new set of coordinates is accordingly obtained. A new iteration takes place when one uses the new set of coordinates to write a new Hamiltonian. The convergence criteria are observed by comparing the variations in the corresponding system's $ \{\eta_{i,j}\} $ and $\{ \psi_{k,i,s} \}$ between consecutive iterations. The procedure can be characterized as an auto-consistent calculation in which the set of coordinates $\{ \eta_{i,j} \}$ is associated to the electronic set $\{ \psi_{k,i,s} \}$ \cite{lima2006dynamical}. 

Crucial to this work is the time evolution of the system. As our model is hybrid, although a coupled one, time dependence of the electronic part is performed differently from that of the lattice. For the electrons, the time-dependent Schr\"odinger equation is used,
\begin{equation}
	|\psi_k(t+dt)\rangle = e^{-\frac i\hslash H(t) dt}|\psi_k(t)\rangle.
\end{equation}
For the classical treatment governing the lattice part of the system, on the other hand, we adopted the Euler-Lagrange equations. Their solutions can be written as a Newtonian equation able to describe the movements of the sites in the system and is given by
\begin{equation}
		M\ddot \eta_{i,j} = \frac12K\left(\eta_{i,j'}+\eta_{i,j''}+\eta_{j,i'}+\eta_{j,i''} \right )-2K\eta_{i,j}+ \frac12\alpha\left(B_{i,j'}+B_{i,j''}+B_{j,i'}+B_{j,i''}-4B_{i,j} + \mathrm{c.c.}\right ),
\end{equation}
where $B_{i,j} \equiv \sum^{'}_{k,s} \psi_{k,s}^*(i,t) \psi_{k,s}(j,t)$ is the term that couples the electronic and the lattice degrees of freedom, and the prime means the summation is over occupied one-particle states.

Another crucial aspect of the model is the presence of the external electric field $\mathrm{\textbf{E}}(t)$. This field is included by inserting a time-dependent vector potential, $\mathrm{\textbf{A}}(t)$, through a Peierls Substitution for the electronic transfer integrals of the system \cite{da2019quasiparticle}, making the hopping term
\begin{equation}
	t_{i,j} = e^{-i\gamma\mathrm{\textbf{A}}}\left(t_0 - \alpha \eta_{i,j} \right ),
\end{equation}
where $ \gamma \equiv ea/(\hslash c) $, with $a$ being the lattice parameter ($ a = 1.42 $ \r{A} in graphene nanoribbons), $e$ being the absolute value of the electronic charge, and $c$ the speed of light. The vector potential relates to the electric field according to $ \mathrm{\textbf{E}}(t) = -(1/c)\dot{\mathrm{\textbf{A}}}(t)$. The electric field is turned on adiabatically so as to avoid artificial numerical oscillations~\cite{ribeiro2013impurity}.

\section{Results}

\begin{figure*}[pos=ht]
	\centering
	\includegraphics[width=0.95\linewidth]{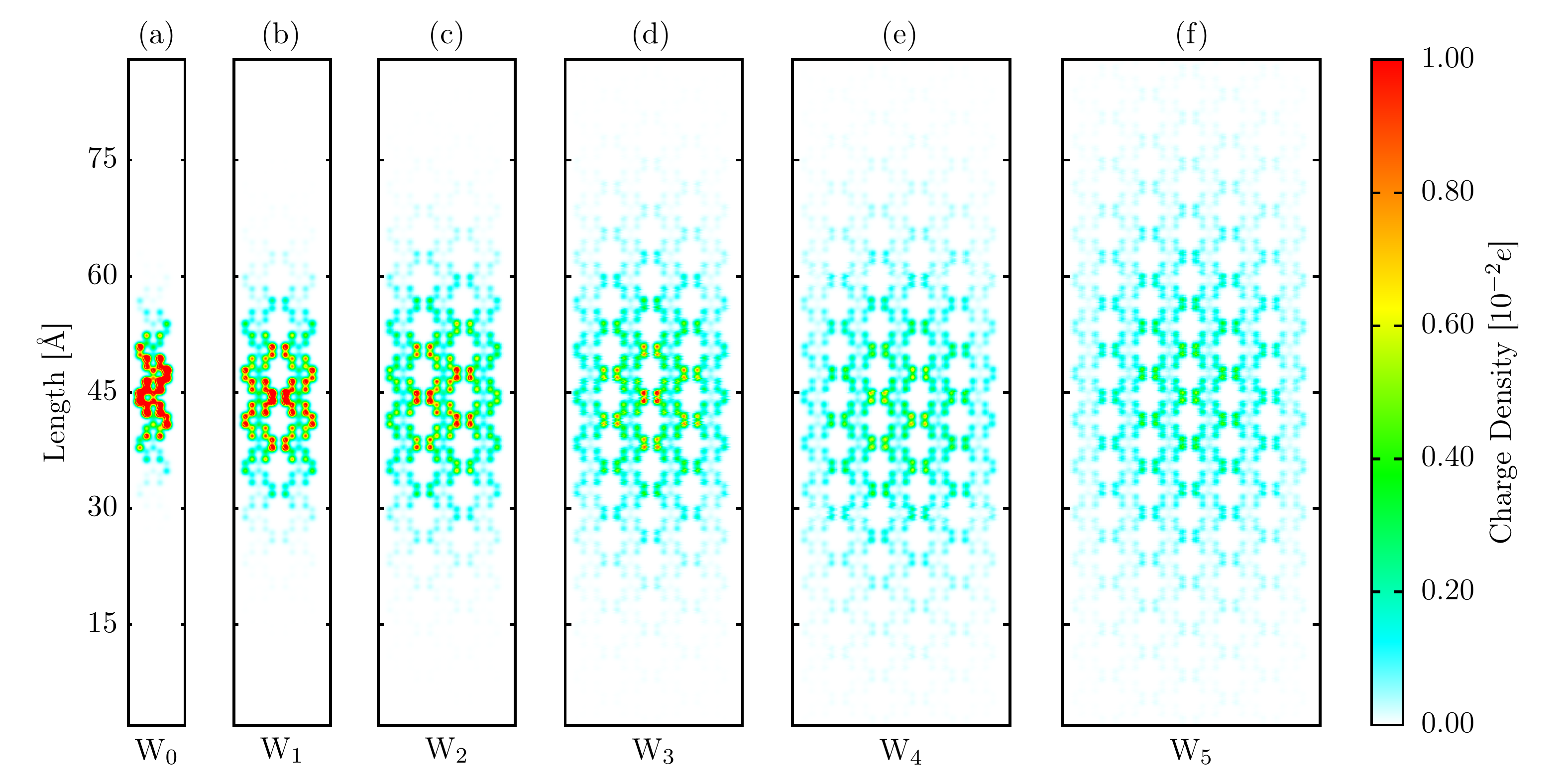}
	\caption{Ground state solutions for the charge carrier distribution on porous nanoribbons with different widths.}
	\label{fig:statcharge}
\end{figure*}

We begin our discussions by characterizing the charge distribution on porous nanoribbons of different widths, as depicted in Figure \ref{fig:statcharge}. For all the cases, an electron is extracted from the chain, thus obtained a net $+1e$ charged system \cite{da2019quasiparticle}. The first feature to be noted is that as the nanoribbon gets broader, the charge is less concentrated. In Figure \ref{fig:statcharge}, lattice regions with high charge concentrations are represented by the red spots. Although the overall trend is in accordance with the expected metallic behavior graphene sheets present \cite{garcia2011group,singh2011graphene}, one can note that even for large widths the semiconducting character of the system, i.e. the carrier confinement trend, is preserved. This property can be attributed to the porosity of the system, as previous work with regular graphene nanoribbons have shown that the semiconductivity is a loss for large enough nanoribbons \cite{da2016polaron}. As a consequence, quasiparticles such as polarons tend to move substantially faster with the increase of the width \cite{da2016polaron,ribeiro2015transport}. Moreover, the charge localization is much smaller than the one previously reported in regular graphene nanoribbons of similar sizes \cite{ribeiro2015transport,silva2018influence}. These are solid indications that the nature of the charge carrier is not of a polaron, mostly for wider porous nanoribbons.

\begin{figure*}[pos=ht]
\centering
\includegraphics[width=0.95\linewidth]{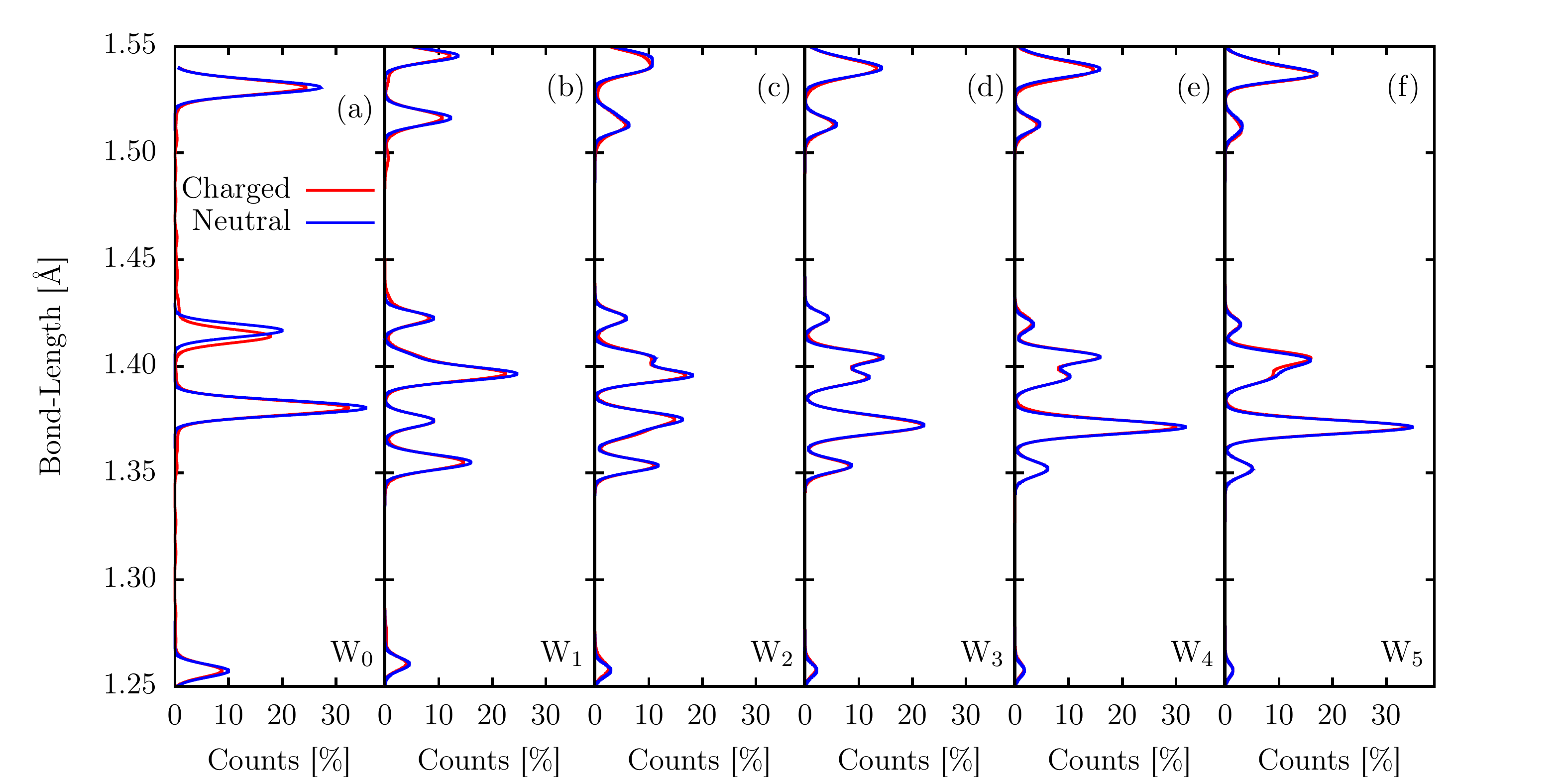}
\caption{Bond-length distribution along the entire length of all the neutral and charged PG nanoribbons studied here.}
\label{fig:bl}
\end{figure*}

Figure \ref{fig:bl} shows the distribution of C-C bond-lengths for the cases presented in Figure \ref{fig:statcharge}. One can note that the profiles for the charged systems are rather similar to that of the neutral ones, as can be noted in Figures \ref{fig:statcharge} (a-f). This is particularly true for broader nanoribbons. A feature that can be inferred from Figures \ref{fig:bl}(d-f) is that the bond-length configuration is not substantially altered from the width W3. The exception is presented in Figure \ref{fig:bl}(a). In this case, the system resembles much more a regular graphene nanoribbon than a PG one. In broader cases, the peak around 1.37 gets larger for wider nanoribbons indicating a convergence trend for the bond-length values when the system changes from nanoribbons towards a sheet with a high surface area. Again, this trend reinforces the argument that a somewhat different scenario from that of polarons takes place in these PG nanoribbons. Differently from polaron endowed regular graphene nanoribbons, one can see in Figures \ref{fig:bl}(b-f) that there is no substantial lattice distortions in the presence of charge when it comes to wider PG nanoribbons. 

\begin{figure*}[pos=ht]
\centering
\includegraphics[width=0.95\linewidth]{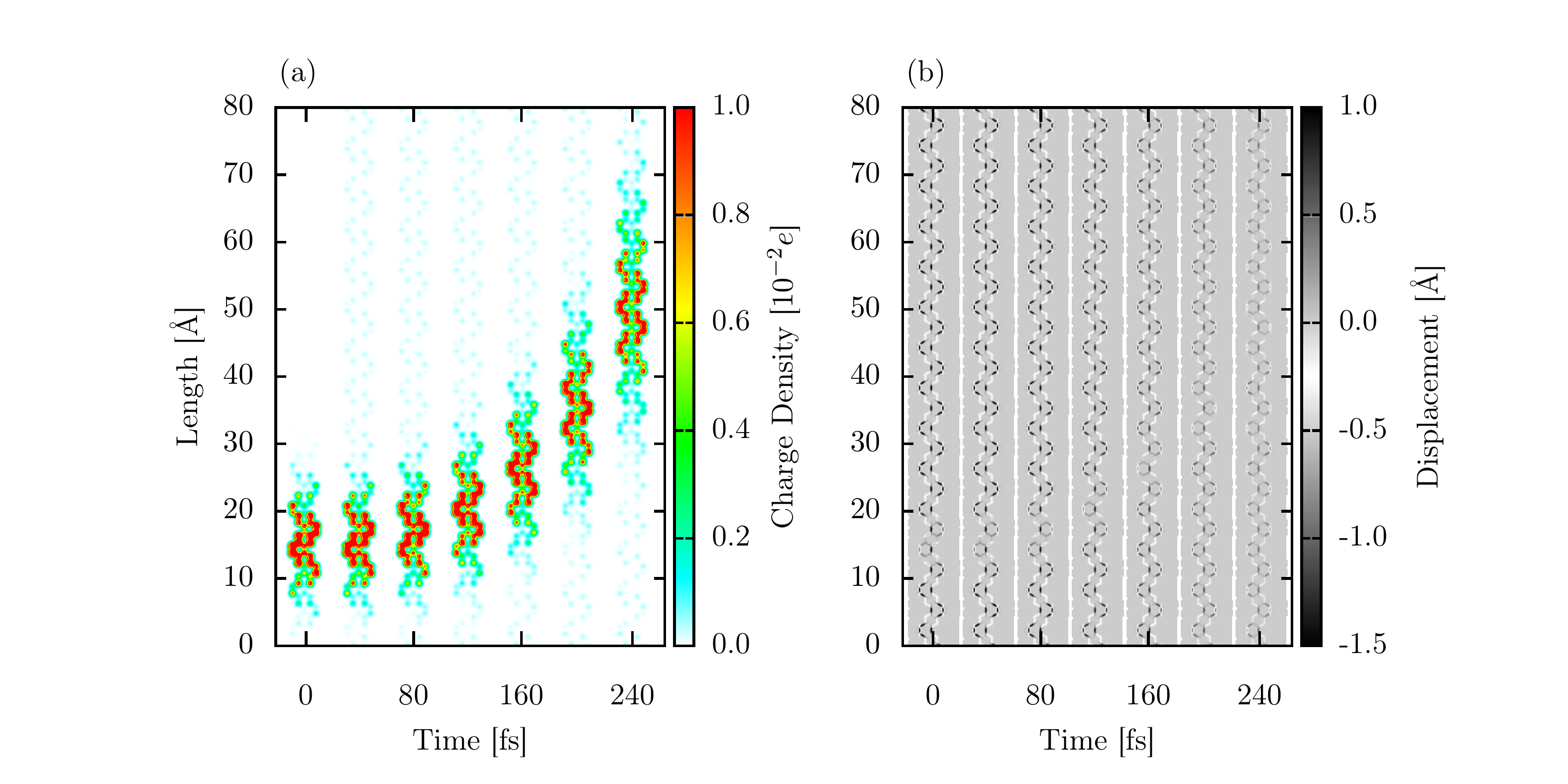}
\caption{Time evolution of the (a) charge density and (b) lattice distortion, of the system corresponding to W0 case.}
\label{fig:w0dyn}
\end{figure*}

In Figure \ref{fig:w0dyn} we include the external electric field and simulates the charge dynamics of the W0-PG system, as a representative case. Figure \ref{fig:w0dyn}(a) presents the charge density time evolution, whereas Figure \ref{fig:w0dyn}(b) shows the time evolution of the its related bond-lengths. One can see that while the charge is accelerated by the electric field, there is no corresponding distortion following the center of the charge. This is a definitive demonstration that the transport mechanism is not mediated by polarons, as is the case in pristine armchair graphene nanoribbons \cite{da2016polaron,ribeiro2015transport}. This is so because polarons are quasiparticles defined as the collective behavior that arises from the correlation between the charge and the phonon excitations. We noted that for the transport mechanism in PG nanoribbons, even in the narrower case studied here, there is a loss of energy related to the lattice conjugation that is transferred to quasiparticle. In Figure \ref{fig:w0dyn}(b), one can realize that the well-distinguished black spots at the beginning of the simulation (high distortions) tend to the grey one at 240 fs, which represents a smaller degree of bond distortion. This energy transfer process is mediated by the electron-phonon coupling term and causes the rapid acceleration of the carrier, as shown in Figure \ref{fig:w0dyn}(a). It is worthwhile to stress that this transport mechanism is not observed in the case of regular graphene nanoribbons \cite{da2016polaron,ribeiro2015transport,da2019quasiparticle}. The results presented in Figure \ref{fig:w0dyn} suggest that charge density waves are responsible for mediating the charge transport mechanism in PG nanoribbons.

\begin{figure*}[pos=ht]
\centering
\includegraphics[width=0.95\linewidth]{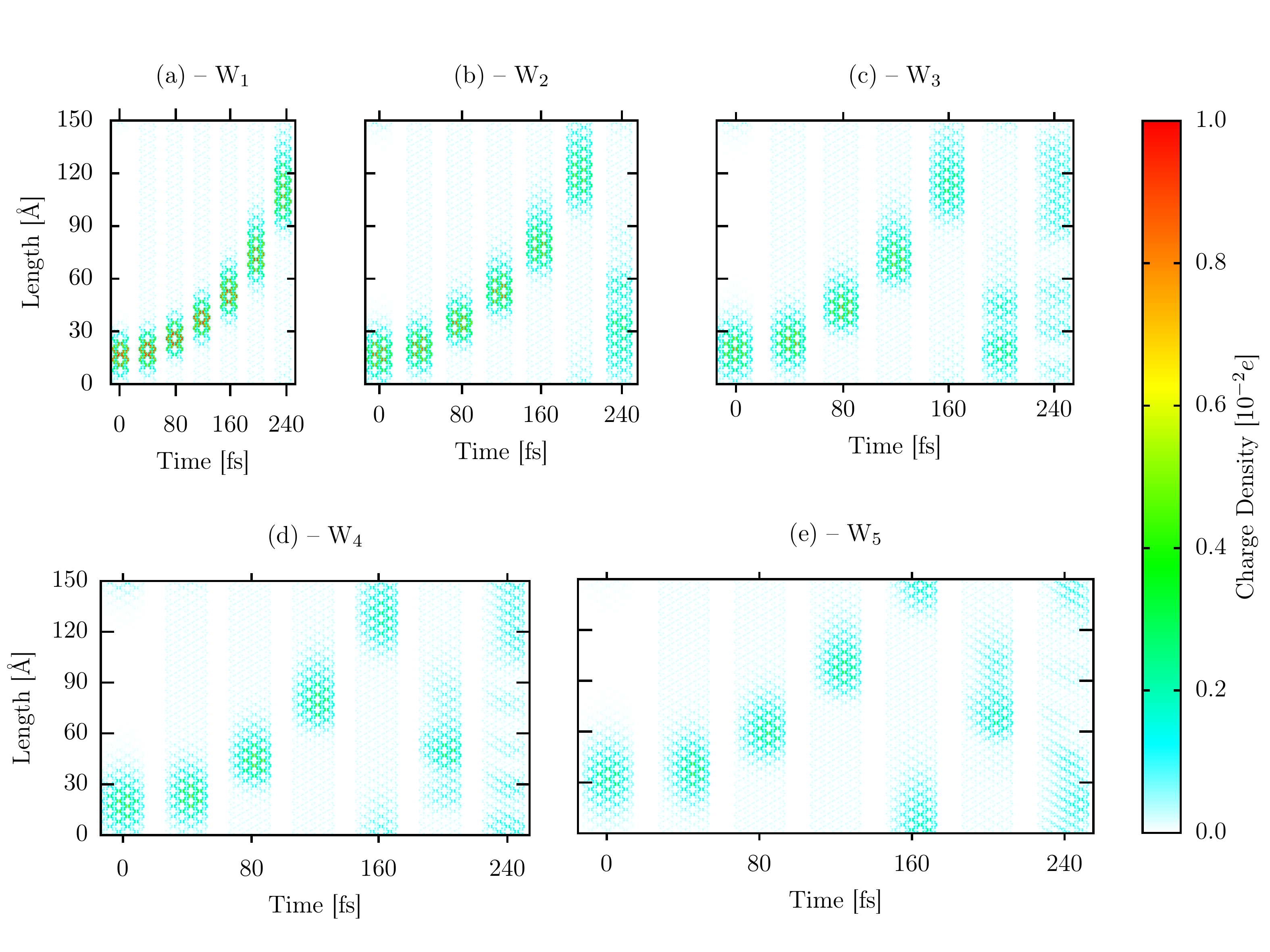}
\caption{Charge density time evolution of (a) W$_1$, (b) W$_2$, (c) W$_3$, (d) W$_4$, and (e) W$_5$ PG nanoribbons.}
\label{fig:dynall}
\end{figure*}

We conducted the same simulations for all the other widths and the same qualitative behavior is perceived, as shown in Figure \ref{fig:dynall}. As the lattice displacement profile is similar to that of the previous figure, in Figure \ref{fig:dynall} we only show the charge density time evolution. The dynamics are different depending on the width of the system, as should be expected from the different structures of the initial state of Figure \ref{fig:statcharge}. As a rule of thumb, the broader the nanoribbon, the larger is the charge carrier mobility. Despite the already established difference in the transport mechanism, this is a feature that is also observed in other graphene nanoribbons \cite{silva2018influence}. The charge density waves move in PG within the optical regime with terminal velocities varying from 0.5 up to 1.15 \r{A}/fs, as can be inferred from Figures \ref{fig:dynall}(a-e). These velocities are lower than the ones for polarons in conventional graphene nanoribbons (2.2-5.1 \r{A}/fs) \cite{da2016polaron,ribeiro2015transport,da2019quasiparticle,silva2018influence}. As the electric field is turned on, although the lattice distortion does not follow the external field --- as it would in a polaronic scenario --- the charge density wave moves, thus transporting the charge. In the wider cases, the charge density waves reach the end of the lattice very fast, thus returning to its beginning due to the periodic boundary conditions (see Figures \ref{fig:dynall}(b-e)). Since the charge movement may disturb the lattice generating phonons, the charge density waves encounter phonons produced by their own movement after crossing the periodic boundary conditions. This phenomenon introduced by boundary conditions choice affects the carriers producing the blurred pattern at 240 fs as presented in Figures \ref{fig:dynall}(b-e).     

\begin{figure*}[pos=ht]
\centering
\includegraphics[width=0.95\linewidth]{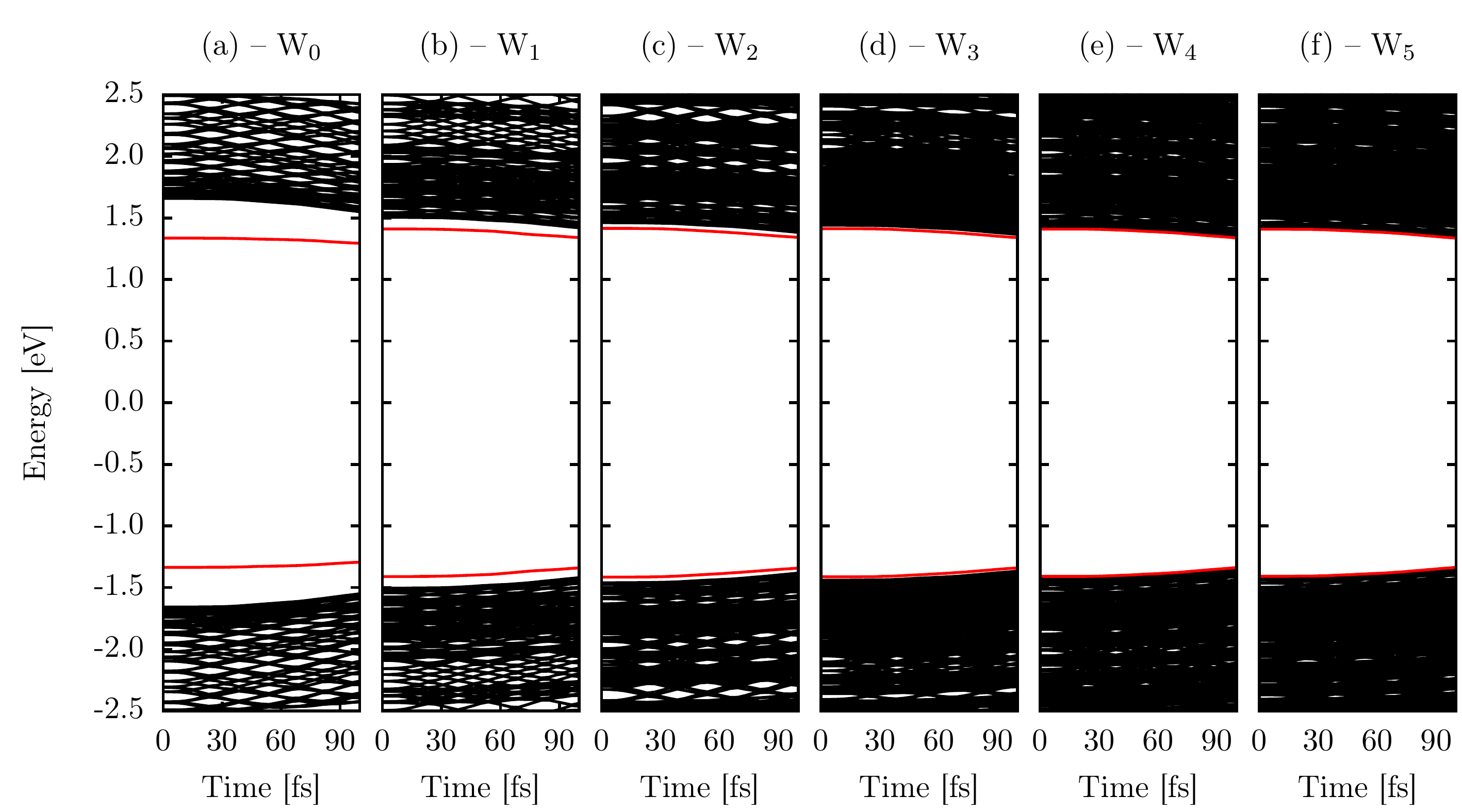}
\caption{Time evolution of the band structure of (a) W0, (b) W$_1$, (c) W$_2$, (d) W$_3$, (e) W$_4$, and (f) W$_5$ PG nanoribbons.}
\label{fig:bands}
\end{figure*}

Finally, Figure \ref{fig:bands} presents the time evolution of the band structure for the charged PG nanoribbons studied here. One can note the presence of energy levels inside the bandgap (red lines). These are the levels that characterizes the charge density wave states. Note that as the systems get broader, the charge density waves levels approaches the conducting and valence band. As the system gets wider a more two-dimensional character prevails, thus explaining the fact that the intragap levels approach the bands. Even so, due to the width limitation of the PG nanoribbons, a semiconducting bandgap still takes place for the wider lattices.

\section{Conclusions}

In summary, we described charge density waves transport mechanism in porous graphene nanoribbons of different widths. So far, the literature had reported quasiparticles such as polarons and bipolarons to be responsible for the transport in graphene-based nanoribbons \cite{pereira2020charge,junior2020charge,pereira2020charge,junior2020transport,silva2019geraldo}. By considering porosity, the one dimensional character that is typical of systems which exhibits such phenomena is present. It was also found that porosity works towards contributing to maintain the semiconductor nature of the nanoribbons, as we have seen that even broad nanoribbons possess a considerable finite bandgap. We noted that for the transport mechanism in PG nanoribbons, even in the narrower case studied here, there is a loss of energy related to the lattice conjugation that is transferred to the charge carrier. This process is mediated by the electron-phonon coupling term and causes the rapid acceleration of the carrier. Importantly, this transport mechanism is not observed in the case of regular graphene nanoribbons \cite{da2016polaron,ribeiro2015transport,da2019quasiparticle}. In PG nanoribbons, the charge is accelerated by the electric field and there is no corresponding distortion following the center of charge. These results suggest that charge density waves are responsible for mediating the charge transport mechanism in PG nanoribbons. Moreover, they are a definitive demonstration that the transport mechanism in PG nanoribbons is not mediated by polarons, as is the case in pristine armchair graphene nanoribbons. As a rule, the broader the nanoribbon, the larger is the charge carrier mobility. The charge density waves move in PG within the optical regime with terminal velocities varying from 0.5 \r{A}/fs up to 1.15 \r{A}/fs. These velocities are lower than the ones for polarons in regular graphene nanoribbons (2.2-5.1 \r{A}/fs) \cite{silva2018influence,ribeiro2015transport,da2016polaron}. 

\section*{Acknowledgements}

The authors gratefully acknowledge the financial support from Brazilian Research Councils CNPq, CAPES, and FAPDF and CENAPAD-SP for providing the computational facilities. M.L.P.J. gratefully acknowledge the financial support from CAPES grant 88882.383674/2019-01. L.A.R.J. and W.F.G. gratefully acknowledge the financial support from FAP-DF grant 0193.0000248/2019-32. L.A.R.J. and G.M.S gratefully acknowledge, respectively, the financial support from CNPq grants 302236/2018-0 and 304637/2018-1. R.T.S.J. gratefully acknowledge, respectively, the financial support from CNPq grants 465741/2014-2 and 312180/2019-5, CAPES grant 88887.144009/2017-00, and FAP-DF grants 0193.001366/2016 and 0193.001365/2016. L.A.R.J. gratefully acknowledges the financial support from DPI/DIRPE/UnB (Edital DPI/DPG 03/2020) grant 23106.057541/2020-89 and from IFD/UnB (Edital 01/2020) grant 23106.090790/2020-86.

\printcredits

\bibliographystyle{unsrt}

\bibliography{cas-refs}

\end{document}